\documentclass[11pt,letter]{article}

\usepackage{amsmath}
\usepackage{amsthm}
\usepackage{xspace}
\usepackage{amssymb}
\usepackage{epsfig}
\usepackage{graphicx}
\usepackage{bbold}
\usepackage[usenames,dvipsnames]{color}
\usepackage{cite}
\usepackage{algorithmic}
\usepackage{algorithm}

\newtheorem{Thm}{Theorem}[section]
\newtheorem{Lem}[Thm]{Lemma}

\newtheorem{Obs}{Observation}

\begin{document}

\title{A Formal Model of Anonymous Systems\footnote{Some proofs are in the appendix.}}

\author{Yang D. Li \footnote{Email: danielliy@gmail.com.  Department of Computer Science and Engineering, The Chinese University of Hong Kong.)}}

\maketitle
\abstract{We put forward a formal model of anonymous systems. And we concentrate on the anonymous failure detectors in our model. In particular, we give three examples of anonymous failure detectors and show that
\begin{itemize}
\item they can be used to solve the consensus problem;

\item they are equivalent to their classic counterparts.
\end{itemize}

Moreover, we show some relationship among them and provide a simple classification of anonymous failure detectors.
}

\newpage

\section{Introduction}

\subsection{Background}

The \emph{consensus} problem \cite{CHT1996} is now recognized as one of the most important problems to solve when one has to
design or to implement reliable applications on top of an unreliable asynchronous distributed system. As it is impossible to implement consensus even with one faulty process \cite{FLP1985}, one of the solutions to this concern is to turn to the concept of \emph{failure detectors}. In \cite{CT1996}, the concept of unreliable failure detectors is introduced and used
to solve the consensus problem in asynchronous systems. In \cite{DFG2002}, the weakest failure detector for solving consensus in the
message-passing model is proved to be $\Omega$ and $\diamond W$ with a majority of correct processes. In \cite{LH1994}, in the shared-memory model, $\Omega$ and $\diamond W$ are the weakest failure detector for solving consensus in any environment. The difference between \cite{DFG2002} and \cite{LH1994} is that \cite{LH1994} has a stronger abstraction, i.e. register, in the process of implementing the consensus problem.

Further, in \cite{DFG2010} and \cite{DFGHKT2004}, a new kind of failure detectors, $\Sigma$, is introduced, which can be used to implement register. Consequently,
the weakest failure detector for solving consensus is actually $(\Omega, \Sigma)$. In \cite{DFG2002}, the realistic failure detectors are
considered, in \cite{MR1999}, the generic protocol for solving consensus is brought forward, and in \cite{Zie2007}, the eventual failure detectors
are classified. Particularly, in \cite{JT2007}, the fact that every problem has a weakest failure detector is shown.

\cite{BR2010} studies the failure detectors in an anonymous system, where the processes have no identity. Nevertheless, it does not provide a mathematical characterization of \emph{anonymity}, the central concept in the paper, which results in the vagueness of the anonymous system. In this paper, we address the question of anonymity. Specifically, we provide a rigorous model for anonymous systems and show several results in our model.

\subsection{A Formal Model for Anonymous Systems}

We use $\mathbb{N}$, the set of natural numbers $\{0, 1, 2, \ldots \}$, to denote the range of the clock's ticks. $P$ means a set of $n$ processes $\{p_1, p_2, \ldots, p_n\}$. $F: \mathbb{N} \rightarrow 2^P$ is a failure pattern, i.e. the set of processes crashed at a certain time. An environment $\mathcal{E}$ is a set of possible failure patterns. The processes can only fail by crashing (halting permanently). We assume that at least one process is correct in our model. Each process is connected to every other process via a reliable channel and message delays on these channels are unbounded but finite.

Communication can be based on the \emph{broadcast} primitive (the same as in the classical system) and an \emph{anonymous receive} operation (to be introduced later). We characterize the anonymity of the system by using a permutation function. Suppose that $\Pi$ is a permutation function mapping from $P$ to $P$. That is to say, $\Pi$ is a permutation of all the processes. Let $\mathcal{R}$ be a possibly infinite range of values that are sent to each other by the processes, $R$ be a function mapping $P \times \mathbb{N}$ to $\mathcal{R}$, and $R_i(j,t)$ be the value that process $i$ receives from another process $j$ at time $t$. Intuitively, the anonymous receive operation means that the receiver cannot tell who sends out the message. Mathematically, the anonymous receive operation is defined by a $R^\Pi$ function:

$$
R^\Pi_i (j,t) = R_i(\Pi(j), t).
$$

The system above is called an \emph{anonymous} \emph{message-passing} system. The anonymous shared-memory system is introduced in the appendix.

\subsection{A Formal Definition of Anonymous Failure Detectors}

Based on our formal model of anonymous systems, we introduce the formal definition of \emph{anonymous failure detectors}. We define $crashed(F)$ (faulty processes) to be $\cup_{t \in \mathbb{N}}F(t)$ and $correct(F)$ (correct processes) to be $P - crashed(F)$. Moreover, $|crashed(F)|$ means the number of faulty processes and $|correct(F)|$ represents the number of correct processes.

A failure detector history $H$ with range $V$ is a function from $P \times \mathbb{N}$ to $V$. $H(p, t)$ is the value of the failure detector module of process $p$ at time $t$. A failure detector $\mathcal{D}$ is a function that maps each failure pattern $F$ to a set of failure detector histories $H_{\mathcal{D}}$ with range $V_\mathcal{D}$ (where $V_\mathcal{D}$ denotes the range of failure detector outputs of $\mathcal{D}$). $\mathcal{D}(F)$ denotes the set of possible failure detector histories permitted by $\mathcal{D}$ for the failure pattern $F$. We define $F^\Pi(t) = \Pi(F(t)), \forall t \in \mathbb{N}$ and $H^\Pi(p, t) = H(\Pi(p), t), \forall p \in P, t \in \mathbb{N}$. If $\forall \Pi, H^\Pi \in \mathcal{D}(F^\Pi)$, then $\mathcal{D}$ is called a \emph{anonymous failure detector}.

\subsection{Examples of Anonymous Failure Detectors}

In this part we introduce some examples of failure detectors under our model of anonymous systems.

\subsubsection{$\mathcal{N}$}
Each failure detector module of $\mathcal{N}$ outputs a natural number in $\{0, 1, 2, \ldots, n\}$, which represents the number of processes suspected to have crashed. So the range of $\mathcal{N}$ is $V_{\mathcal{N}} = \{0, 1, 2, \ldots, n\}$.

$\mathcal{N}(F)$ is the set of all failure detector histories $H_\mathcal{N}$ with range $V_{\mathcal{N}}$ that satisfies the following properties:

\begin{itemize}
\item Completeness : Eventually the failure detector outputs a number that is greater than or equal to the actual number of
crashed processes.

$\exists t \in \mathbb{N}, \forall t' \in \mathbb{N} \text{ and } t' \ge t , \forall q \in P, H_{\mathcal{N}}(q, t') \ge |crashed(F)|$.

\item Accuracy : The (correct) failure detector always outputs a number that is smaller than or equal to the actual number of crashed
processes.

$\forall t \in \mathbb{N}, \forall q \in correct(F), H_{\mathcal{N}}(q, t) \le |crashed(F)|$.

\end{itemize}

\subsubsection{$\diamond \mathcal{N}$}

Each failure detector module of $\diamond \mathcal{N}$ outputs a natural number in $\{0, 1, 2, \ldots, n\}$, which represents the number of processes suspected to
have crashed. So the range of $\mathcal{N}$ is $V_{\diamond \mathcal{N}} = \{0, 1, 2, \ldots, n\}$.

$\diamond \mathcal{N}(F)$ is the set of all failure detector histories $H_{\diamond \mathcal{N}}$ with range $V_{\diamond \mathcal{N}}$ that satisfies the following properties:

\begin{itemize}

\item Completeness : Eventually the failure detector outputs a number that is greater than or equal to the actual number of
crashed processes.

$\exists t \in \mathbb{N}, \forall  t' \in \mathbb{N}\text{ and }t' \ge t , \forall q \in P, H_{\diamond \mathcal{N}}(q, t') \ge |crashed(F)|$.

\item Eventual Accuracy : Eventually, the output of (correct) failure detectors is a number that is smaller than or equal to the actual number of crashed
processes.

$\exists t \in \mathbb{N}, \forall t' \in \mathbb{N} \text{ and } t' > t , \forall q \in correct(F), H_{\diamond \mathcal{N}}(q, t') \le |crashed(F)|$.

\end{itemize}

\subsubsection{$\Theta$}

Each failure detector module of $\Theta$ outputs a boolean value, i.e., true or false. So under this circumstance, the range of $\Theta$ is $V_{\Theta} = \{true, false\}$.

$\Theta(F)$ is the set of all failure detector histories $H_{\Theta}$ with range $V_{\Theta}$ that satisfies the following property:

\begin{itemize}
\item Eventual Self-Trust: There is a time after which there is only one correct process, which trusts itself.

$\exists t \in \mathbb{N}, \exists p \in correct(F)$, s.t. $\forall t' \in \mathbb{N}\text{ and }t' > t, H(t', p) = true, \forall q \in correct(F) - p, H_{\Theta}(t', q) = false.$

\end{itemize}

\section{Consensus Algorithms}

\subsection{Consensus Problem}

In this part we briefly review the consensus problem \cite{CT1996}, which is defined by the following four properties:

\begin{itemize}
\item Termination: Every correct process eventually decides some value.

\item Irrevocability (Integrity): Every process decides at most once.

\item Agreement: No two correct processes decide differently.

\item Validity: If a process decides $v$, then $v$ was proposed by some process.

\end{itemize}

\subsection{Consensus Algorithm with $\mathcal{N}$}

We assume that the number of processes that may crash is bounded by $f$. Our $\mathcal{N}$ based consensus algorithm proceeds in $f+1$ asynchronous rounds. In each round every process broadcasts its value. Then it blocks until it has received enough round-$r$ messages. In this and the following sections we assume that the message system places all received messages in an multi-set called \emph{received}. Since we only consider messages from the round a process is currently in, we implicitly delay messages from later rounds until their round starts. For memory efficiency, messages from previous rounds can be discarded at every round switch (i.e., whenever $r$ is increased).

{\bf Remark: } Since algorithms typically wait for messages from alive processes, we deemed it more useful to use the converse of the failure detector described above. Moreover, since it is always safe to wait for messages from $n - f$ processes, we consider oracles that output the number of processes believed to be alive, denoted by $\mathcal{N}$ and $\diamond \mathcal{N}$ respectively. In the following, we will use $\mathcal{N}$ and $\diamond \mathcal{N}$ to denote the output of $\mathcal{N}$ and $\diamond \mathcal{N}$ respectively.

\begin{algorithm}
\caption{Consensus on $v$ with $\mathcal{N}$}
\label{Algorithm $1$}
\begin{algorithmic}[1]
\STATE $v \in \{0, 1\}$ initially the input value
\FOR{$r$ from $1$ to $f + 1$}
\STATE broadcast $(Propose, v, r)$
\STATE wait until received contains $(Propose, v, r)$ at least $\mathcal{N}$ times
\STATE $values \leftarrow  \{v\} \cup \{v' : (Propose, v', r) \in received\}$
\STATE $v \leftarrow  \max v' \in values$
\ENDFOR
\STATE decide $v$
\end{algorithmic}
\end{algorithm}

In this part, we show that consensus is solvable among $n \ge f + 1$ processes. To this end we start out with a lemma that shows that processes will never give up on the value $1$ once they have adopted it.

\begin{Lem}[Stubbornness]\label{lemma1.1}
If some correct process $p$ adopts $v \leftarrow 1$ in some round $r$ or $p$ initially ($r = 0$) proposes $1$, then $p$ will have $v = 1$ for all rounds $r' > r$.
\end{Lem}

With this intermediate step, showing Validity becomes quite simple:

\begin{Lem}[Validity]\label{lemma1.2}
If a process decides $v$ using Algorithm $1$, then $v$ was proposed by some process.
\end{Lem}

The proof of agreement is patterned around the idea, that among the $f + 1$ rounds there must be one round during which no process crashes. We will show that this is enough for all
processes to reach states such that all preferred values are equal and that once this is the case, no process can decide on another value, since no variable or message will ever carry the other value in later rounds.

\begin{Lem}\label{lemma1.3}
If there exists one round, say $r$, which no process is in when crashing all processes will set their v to the same $\max(values)$ by the end of that round, and decide on this $v$.
\end{Lem}

Observing that there are $f+1$ rounds but at most $f$ processes can crash, it is evident that:

\begin{Obs}\label{observation1.1}
In executions with $f+1$ rounds, where at most $f$ processes can crash there is at least one round in which no processes crash.
\end{Obs}

Agreement is evident from the Lemma \ref{lemma1.3} and Observation \ref{observation1.1}. Validity was shown in Lemma \ref{lemma1.2}. Irrevocability follows trivially from the algorithm, and Termination follows from the fact that the algorithm can never get stuck in a round, since the number of received messages must eventually be greater or equal to the output of $\mathcal{N}$ in every round.

\begin{Thm}
Algorithm $1$ allows $n > f$ processes to reach Consensus.
\end{Thm}

\subsection{Consensus Algorithm with $\diamond \mathcal{N}$}

We show that consensus is possible among $n > 2f$ processes when we augment our basic asynchronous model with $\diamond \mathcal{N}$.

\begin{algorithm}
\caption{Consensus algorithm with $\diamond \mathcal{N}$}
\label{Algorithm $2$}
\begin{algorithmic}[1]
\STATE $v \in \{0, 1\}$ initially the input value
\STATE $lock \leftarrow ?$, $decided \leftarrow false$, $r \leftarrow 0$
\LOOP
\STATE broadcast $(Propose, r, v)$
\STATE wait until received contains $(Propose, r, \_)$ at least $\diamond \mathcal{N}$ times
\STATE proposed   $\{w : (Propose, r,w) \in received\}$
\STATE $v \leftarrow  \min(proposed)$
\IF {$proposed - \{v\} = \emptyset$}
\STATE $lock \leftarrow v$
\ELSE
\STATE $lock \leftarrow ?$
\ENDIF
\STATE broadcast $(Lock, r, lock, v)$
\IF {decided}
\STATE halt
\ENDIF
\STATE wait until received contains $(Lock, r, \_, \_)$ at least $\diamond \mathcal{N}$ times
\STATE $locked  \leftarrow \{w : (Lock, r,w, ) \in received\}$
\IF {$locked - \{?\} \ne \emptyset$}
\STATE $v \leftarrow  \min(locked - \{?\})$
\IF {$locked - {v} = \emptyset$}
\STATE decide $v$
\STATE $decided \leftarrow  true$
\ENDIF
\ELSE
\STATE $proposed \leftarrow \{w : (Lock, r, \_, w) \in received\}$
\STATE $v \leftarrow  \min(proposed)$
\ENDIF
\STATE $r \leftarrow  r + 1$
\ENDLOOP
\end{algorithmic}
\end{algorithm}

\begin{Lem}\label{lemma2.1}
When some process $p$ sends a lock message for some value, say $x \ne ?$, in round $r$, then no other process can send a lock message for some other value $y \notin \{x, ?\}$.
\end{Lem}

\begin{Lem}\label{lemma2.2}
In Algorithm $2$, let round $r$ be the first round where some process decides, say on $x$, then (1) no other process can decide a different value in round $r$, and (2) all other processes
will decide $x$ at most one round later.
\end{Lem}

What remains to be shown is that there will eventually be a round in which one process is able to decide.

\begin{Lem}\label{lemma2.3}
In every execution of Algorithm $2$ there eventually is a round $r_d$ where at least one process decides.
\end{Lem}

\begin{Lem}\label{lemma2.4}
Algorithm $2$ guarantees Validity and Integrity.
\end{Lem}

\begin{Thm}
Algorithm $2$ solves consensus in anonymous asynchronous systems augmented with $\diamond N$ when $n > 2f$.
\end{Thm}

\subsection{Consensus Algorithm with $\Theta$}

\begin{algorithm}
\caption{Consensus algorithm with $\Theta$}
\label{Algorithm $3$}
\begin{algorithmic}[1]
\STATE $v \in \{0, 1\}$ initially the input value
\STATE $r \leftarrow 0$

\STATE Code for processes $p$:
\LOOP
\STATE wait until $\Theta = true$ or $received$ contains $(Leader, r, \_)$ at least once
\IF {$received(Leader, r, w)$}
\STATE $v \leftarrow w$
\ELSE
\IF {\textbf{if} $ \Theta = true$}
\STATE $broadcast (Leader, r, v)$
\ENDIF
\ENDIF
\STATE $broadcast (Report, r, v)$
\STATE wait until $received$ contains $(Report, r, \_)$ at least $(n - f)$ times
\IF {$\exists w : received(Report, r, \_)$ from $> n/2$ processes}
\STATE $aux \leftarrow w$
\ELSE
\STATE $aux \leftarrow ?$
\ENDIF
\STATE broadcast$(Vote, r, aux)$
\STATE wait until $received$ contains $(Vote, r, \_)$ at least $(n - f)$ times
\IF {$received(Vote, r, aux')$ with $aux' \ne ? $}
\STATE $v \leftarrow  aux'$
\ENDIF
\IF {$received(Vote, r, aux')$ with $aux' \ne ?$ at least $n - f$ times}
\STATE broadcast$(Decide, v)$
\ENDIF
\STATE $r \leftarrow  r + 1$
\ENDLOOP

\STATE upon reception of $(Decide, v)$ do
\STATE broadcast$(Decide, v)$
\STATE decide $decision$
\STATE halt
\end{algorithmic}
\end{algorithm}

The difference between $\Omega$ and $\Theta$ is that with $\Theta$ only the eventual leader learns it¡¯s role directly from the oracle. The most important difference is that in our algorithm only the leader sends a $(Leader, r, \_)$ message. This (and that $n > 2f$) ensures that processes cannot make too much progress independently. First, however, we prove that all processes actually do make progress.

\begin{Lem}\label{lemma3.1}
No correct process blocks forever in a round.
\end{Lem}

Next we show that each process has to decide, by showing that when one or more process decides first then all other processes must decide later on, since deciding is always triggered
by a $(Decide,\_)$ message which processes forward before deciding. Then we prove that there is at least one first process to send that message after the leader has stabilized. Obviously
the first part also holds if the $(Decide,\_)$ message is sent before stabilization.

\begin{Lem}\label{lemma3.2}
Every correct process decides.
\end{Lem}

Through our final Lemma it will become evident that Agreement must hold:

\begin{Lem}\label{lemma3.3}
It is impossible for $(Decide, w)$ and $(Decide, w')$ with $w\ne w'$ to be sent.
\end{Lem}

Since processes decide on the value received via a $(Decide, v)$ message, Agreement follows from the previous Lemma, Termination from Lemma \ref{lemma3.2}, Integrity from the fact that processes halt immediately after deciding and Validity from the fact that all values ever sent, can be easily traced back to an initial value of some process $v$. Therefore we have:

\begin{Thm}
Algorithm $3$ solves Consensus in asynchronous anonymous systems augmented with $\Theta$, if $n > 2f$.
\end{Thm}

\section{Equivalence with Classical Failure Detectors}

In this section we investigate the relationship between our anonymous failure detectors and the classic ones $(\mathcal{P}, \diamond \mathcal{P}$ and $\Omega)$ \cite{CHT1996}\cite{CT1996}. To this end we have to assume that unique identifiers are available and that every reception can be attributed to the sender.

Firstly, we observe that the equivalence between $\Omega$ and $\Theta$ is obvious: to obtain one from the other it is sufficient for the process that trusts itself to simply tell the other processes, or for all processes but the leader elected by $\Omega$ to simply ignore the failure detectors output. The translation of $\mathcal{P}$ to $\mathcal{N}$ and of $\diamond \mathcal{P}$ to $\diamond \mathcal{N}$ are obvious as well: in both cases it suffices to output the number of processes which are not suspected.

The remaining relations are explored in more detail via transformations. By $\mathcal{DF}$ we denote the asynchronous algorithm that implements $\mathcal{F}$ based on $\mathcal{D}$. Both transformations work by building a estimate of the alive processes, denoted by $AL$, and then suspecting all processes that are not in this set, i.e., $P - AL$, where $P$ denotes the set of all processes.

\begin{algorithm}
\caption{$\diamond \mathcal{N} \diamond \mathcal{P}$ Implementation}
\label{Algorithm $4$}
\begin{algorithmic}[1]
\STATE Code for processes $p$:
\STATE $r \leftarrow 0$
\STATE $suspect \leftarrow \emptyset$
\LOOP
\STATE $r \leftarrow r + 1$
\STATE broadcast $(ALIVE, r)$
\STATE wait until received $\diamond \mathcal{N} (ALIVE, r)$ messages from the set $AL$
\STATE $suspect \leftarrow P - AL$
\ENDLOOP
\end{algorithmic}
\end{algorithm}

The implementation of $\diamond \mathcal{N} \diamond \mathcal{P}$ (Algorithm $4$) is quite simple. Since only eventual Strong Accuracy is required, it suffices to output those processes that did not send $(Alive, \_)$ messages in the current round. Thus wrong suspicions can only occur in rounds where the crashed processes have sent messages before crashing, and these messages are faster than those from alive processes. In some later round, this wrong is [eventually] corrected, due to the absence of a message from the crashed process.

Let us now reiterate the properties of $\mathcal{P}$ and $\diamond \mathcal{P}$:

\begin{itemize}

\item Strong Completeness. Eventually ever process that crashes is permanently suspected by every correct process.

\item Strong Accuracy. No process is suspected before it crashes.

\item Eventual Strong Accuracy. There is a time after which correct processes are not suspected by any correct process.

\end{itemize}

\begin{Thm}\label{diamond}
The implementation $\diamond \mathcal{N} \diamond \mathcal{P}$ (Algorithm $4$) guarantees Strong Completeness and Eventual Strong Accuracy.
\end{Thm}

\begin{algorithm}
\caption{$\mathcal{N}\mathcal{P}$ Implementation}
\label{Algorithm $5$}
\begin{algorithmic}[1]
\STATE Code for processes $p$:
\STATE $r \leftarrow 0$
\STATE $suspect \leftarrow \emptyset$; $earlieralive \leftarrow \emptyset$; $lastchange \leftarrow 0$
\LOOP
\STATE broadcast$(ALIVE, r)$
\STATE wait until received $(ALIVE, r)$ messages from some set $AL$ and $|AL| = \mathcal{N}$
\IF {$AL \ne earlieralive$}
\STATE $lastchange \leftarrow r$
\ELSE
\IF {$r \ge lastchange + f + 2$}
\STATE $suspect \leftarrow P - AL$
\ENDIF
\ENDIF
\STATE $earlierlive \leftarrow AL$
\STATE $r = r + 1$
\ENDLOOP
\end{algorithmic}
\end{algorithm}

Since $\mathcal{P}$ is not allowed to make wrong suspicions, we have to make sure that $AL$ always contains all processes that have not crashed whenever we update suspect. To ensure this we wait until this set does not change for $f + 2$ rounds. Before we turn to proving that our translation guarantees Strong Accuracy and Strong Completeness, we show that all alive processes proceed through the rounds in a somewhat coordinated way:

\begin{Lem}\label{lemma4.1}
At any time, the difference between the round numbers of two alive processes $p$ and $q$ is smaller than or equal to $f + 1$.
\end{Lem}

\begin{Lem}\label{lemma4.2}
The Translation $\mathcal{NP}$ guarantees Strong Accuracy.
\end{Lem}

\begin{Lem}\label{lemma4.3}
The Translation $\mathcal{NP}$ guarantees Strong Completeness.
\end{Lem}

From the lemmas above it follows immediately that

\begin{Thm}
The Translation $\mathcal{NP}$ guarantees Strong Completeness and Strong Accuracy.
\end{Thm}

\section{A Simple Classification and Reductions}

\subsection{A Formal Model for Reductions among Anonymous Failure Detectors}

We say an anonymous failure detector $\mathcal{D'}$ can be reduced to another failure detector $\mathcal{D}$ ($\mathcal{D}$ is stronger than $\mathcal{D'}$) if there is an algorithm $T_{\mathcal{D} \rightarrow \mathcal{D'}}$ that transforms $\mathcal{D}$ to $\mathcal{D'}$ under an environment $\mathcal{E}$. $T_{\mathcal{D} \rightarrow \mathcal{D'}}$ (using $\mathcal{D}$) maintains a variable $output_p$ at every process $p$, which emulates the output of $D'$ at $p$. Let $O$ be the history of all the output variables and we require $O^\Pi \in \mathcal{D'}(F^\Pi)$, where $F \in \mathcal{E}$ and $\Pi$ is a permutation function of all the processes.

\subsection{A Simple Classification}

In short, $\mathcal{N}$ and $\diamond \mathcal{N}$ belong to the same class of failure detectors, the \emph{symmetric} failure detectors. $\Theta$  is another class of failure detectors, the \emph{unsymmetrical} failure detectors. The difference of the two kinds of failure detectors is that the symmetric failure detector outputs the same information at all correct processes while the unsymmetrical failure detectors do not.

This is a very simple classification. In an anonymous system, a process cannot distinguish other processes and only knows itself. So usually unsymmetrical failure detectors output one value at a specific process and output some other value at the rest of processes.

\subsection{Reductions among Anonymous Failure Detectors}
Now let¡¯s talk about the relations of the anonymous failure detectors mentioned above.

\begin{Thm}
$\mathcal{N}$ is stronger than $\diamond \mathcal{N}$.
\end{Thm}

\proof
They both have the property of completeness. The accuracy property of $\mathcal{N}$ clearly implies the eventual property of $\diamond \mathcal{N}$ while the reverse is not true.
\endproof

\begin{Thm}\label{incomparable}
$\mathcal{N}$ and $\Theta$ are incomparable.
\end{Thm}

\proof
Obviously, there is no deterministic reduction from $\Theta$ to $\mathcal{N}$ as there is simply no way to break the symmetry in $\mathcal{N}$. Further, there is no deterministic reduction from $\mathcal{N}$ to $\Theta$. By contradiction, assume that there exists a reduction algorithm $A$ such that for each failure pattern $F$ and failure detector history $H \in \Theta(F)$, $A$ outputs a failure detector history $H' \in \mathcal{N}(F)$. Denote $q$ to be a correct process. Since $\mathcal{N}$ can be implemented, there should exists $t_0 \in \mathcal{N}$ such that after $t_0$, i.e. for $t > t_0$, $H' \in \mathcal{N}(F)$ should satisfy completeness and accuracy, i.e. $H_{\mathcal{N}} (q, t) = |crashed(F)|$. Without the loss of generality, suppose that the only process that trusts itself in the $\Theta$ output is process $p_1$. Then we let $p_1$ be silent until $t_1 > t_0$, then at time $t_2$ that $t_0 < t_2 < t_1$, the run of the algorithm cannot distinguish the circumstance that $p_1$ is slow from the one that $p_1$ is dead. Therefore, $H_{\mathcal{N}} (q, t_2) > |crashed(F)|$, which violates the property of accuracy in the definition of $\mathcal{N}$.

\endproof

\begin{Thm}
$\diamond \mathcal{N}$ and $\Theta$ are incomparable.
\end{Thm}
\proof
Following a similar argument in the proof of the previous theorem, it is easy to show that $\diamond \mathcal{N}$ cannot be reduced to $\Theta$. Also, there is no reduction from $\Theta$ to $\diamond \mathcal{N}$ due to symmetry reasons.
\endproof

\subsection{Randomized Reductions}

The reductions discussed above are \emph{deterministic} reductions. We also define the notion of \emph{randomized} reduction. Instead of requiring $O^\Pi \in \mathcal{D'}(F^\Pi)$, we allow the use of randomness and only require $O^\Pi \in \mathcal{D'}(F^\Pi)$ to be correct with probability at least $2/3$. We show an example of randomized reductions.

\begin{Thm}
Under randomized reduction, $\mathcal{N}$ is stronger than $\Theta$.
\end{Thm}

\proof

We will show a reduction algorithm converting $\mathcal{N}$ to $\Theta$. The converse is not possible due to the proof of Theorem \ref{incomparable}.

At first, for each process $p_i$, it randomly generates a real number. In theory, the number of real numbers are infinite and so the chance for two processes that get the same real number is $0$. However, in practice, it may be hard to generate an infinite number of numbers. So here the pool may be finite and there always exists the probability that two processes get the same real number. However, if we let the pool to be large enough, much larger than the number of processes, then the probability for two processes to get the same real number is extremely low and will tend to $0$ if the pool is going to infinity.

This is where the randomness lies. Then we can assume that each process has a distinct real number. In the following process, when it tries to broadcast, it should include this real number. You may think that we have return to the situation of the classic systems. In some sense this thinking is right and some other sense, it is not. Although $p_1$ receives distinct real numbers, it cannot tell whether a real number it receives, is from $p_2, p_3, \ldots,$ or $p_n$. Thus this is consistent with the definition of the anonymous system model.

Then why do we say that in some sense we just return to the classic systems? This can be attributed to the anonymous model. The anonymity is that our real numbers are a permutation of the process id's, in which sense we cannot distinguish the processes. Nonetheless, if we treat the real numbers as the identifiers, then the anonymity just disappears. Then we can comfortably utilize the CHT proof to extract a certain real number corresponding to a certain process that we do not know. Each process at this stage can judge if the extracted real number is its own initial value. If the answer to this is affirmative, then this process is just the process we seek in the $\Theta$ failure detector.

\endproof

\section{Concluding Remarks}
In summary, we provide a rigorous model for anonymous systems and discuss some issues related to failure detectors under our model. We hope that further problems and notions can be brought forward in our model. For instance, more examples of failure detectors can be shown. Moreover, we believe that the major open problem in the anonymous system is the weakest anonymous failure detector for consensus. It may be hard to know the weakest anonymous failure detector in a deterministic sense; so we have defined randomized reduction. We hope that the weakest anonymous failure detector for consensus under randomized reduction can be easier.

\newpage

\bibliography{anonymous}

\newcommand{\etalchar}[1]{$^{#1}$}
\begin{thebibliography}{DGFG{\etalchar{+}}04}

\bibitem[BR10]{BR2010}
Francois Bonnet and Michel Raynal.
\newblock Anonymous asynchronous systems: The case of failures detectors.
\newblock {\em Proceedings of the 24th International Symposium on Distributed
  Computing}, pages 206--220, 2010.

\bibitem[CHT96]{CHT1996}
Tushar Chandra, Vassos Hadzilacos, and Sam Toueg.
\newblock The weakest failure detector for solving consensus.
\newblock {\em Journal of the ACM}, 43(4):685--722, 1996.

\bibitem[CT96]{CT1996}
Tushar Chandra and Sam Toueg.
\newblock Unreliable failure detectors for reliable distributed systems.
\newblock {\em Journal of the ACM}, 43(2):225--267, 1996.

\bibitem[DGFG02]{DFG2002}
Carole Delporte-Gallet, Hugues Fauconnier, and Rachid Guerraoui.
\newblock A realistic look at failure detectors.
\newblock {\em Proceedings of the 2002 International Conference on Dependable
  Systems and Networks}, pages 345--352, 2002.

\bibitem[DGFG{\etalchar{+}}04]{DFGHKT2004}
Carole Delporte-Gallet, Hugues Fauconnier, Rachid Guerraoui, Vassos Hadzilacos,
  Petr Kouznetsov, and Sam Toueg.
\newblock The weakest failure detectors to solve certain fundamental problems
  in distributed computing.
\newblock {\em Proceedings of the 23rd annual ACM symposium on Principles of
  distributed computing}, pages 338--346, 2004.

\bibitem[DGFG10]{DFG2010}
Carole Delporte-Gallet, Hugues Fauconnier, and Rachid Guerraoui.
\newblock Tight failure detection bounds on atomic object implementations.
\newblock {\em Journal of the ACM}, 57(4):Artical 22, 2010.

\bibitem[FLP85]{FLP1985}
Michael Fischer, Nancy Lynch, and Michael Paterson.
\newblock Impossibility of distributed consensus with one faulty process.
\newblock {\em Journal of the ACM}, 32(2):374--382, 1985.

\bibitem[JT07]{JT2007}
Prasad Jayanti and Sam Toueg.
\newblock Every problem has a weakest failure detector.
\newblock {\em Proceedings of the 27th ACM symposium on Principles of
  distributed computing}, pages 75--84, 2007.

\bibitem[LH94]{LH1994}
Wai-Kau Lo and Vassos Hadzilacos.
\newblock Using failure detectors to solve consensus in asynchronous
  shared-memory systems.
\newblock {\em Proceedings of the 8th International Workshop on Distributed
  Algorithms}, pages 280--295, 1994.

\bibitem[MR99]{MR1999}
Achour Mostefaoui and Michel Raynal.
\newblock Solving consensus using {C}handra-{T}oueg's unreliable failure
  detectors: A general quorum-based approach.
\newblock {\em Proceedings of the 13th International Symposium on Distributed
  Computing}, pages 49--63, 1999.

\bibitem[Zie07]{Zie2007}
Piotr Zielinski.
\newblock Automatic classification of eventual failure detectors.
\newblock {\em Proceedings of the 21st International Symposium on Distributed
  Computing}, pages 465--479, 2007.

\end{thebibliography}
\bibliographystyle{plain}

\newpage

\begin{appendix}

\section{A Formal Model of Anonymous Shared-Memory System}

As in the anonymous message-passing system, there is a set of processes $P = \{p_1, p_2, \ldots, p_n\}$ in the anonymous shared-memory system. In addition, there are $m$ objects $O = \{o_1, o_2, \ldots, o_m\}$. We assume that each process has a (possibly infinite) state machine and a set of states, one of which is the initial state. Each state $q$ of process $p$ has three special fields:

\begin{itemize}
\item $q.obj$, the object to be accessed next, or null
\item $q.op$, the operation on $q.obj$ to be executed
\item $q.in$, the input parameter (if any) of $q.op$
\end{itemize}

We use a permutation function $\Pi$, which maps $P$ to $P$. The configuration $C$ of the system is the states of all processes and the values of all shared objects, i.e. vector $(q_1, q_2, \ldots, q_n, v_1, v_2, \ldots, v_m)$. We define $q_i$ to be the state of $\Pi(p_i)$, for $i = 1, 2, \ldots, n$, and $v_m$ to be the value of $o_m$, for $m = 1, 2, \ldots, m$. The function $f$ is the state transition machine from some state and value $(q, v)$ to some state $q'$.

Therefore, just like what we did in the anonymous message-passing model, we also use permutation to characterize the anonymity in the anonymous shared-memory system.

\section{Proofs for Algorithm $1$}

\subsection{Proof of Lemma \ref{lemma1.1}}

A process $p$ always adds its current value to values in line $5$, and always chooses the maximum of all values in line $6$, therefore once $1$ was adopted it will always remain the maximum
(since $0$ and $1$ are the only possible values) and the Lemma follows.

\subsection{Proof of Lemma \ref{lemma1.2}}

Since we are considering binary consensus only, there are only two cases where Validity could be violated: (1) Either some process $p$ decided 1 when all processes had $0$ as their
initial value, or (2) some process $p$ decided $0$ and all processes had $1$ as their initial value. In case (1) $p$ must have received $1$ from some other process at some point, otherwise
$1$ cannot become a member of values, and $p$ initially proposes $0$ by assumption. Since all processes only send their current estimate $v$, some process must have initially proposed $1$,
which is a contradiction to the assumption of (1). The impossibility of case (2) follows from Lemma \ref{lemma1.1}.

\subsection{Proof of Lemma \ref{lemma1.3}}

Let $f_{r-1}$ denote the number of processes that have crashed up to and including round $r-1$. When no processes crash during round $r$, the processes will wait for messages from
$n - f_{r-1}$ processes, since $\mathcal{N}$ will never output a number smaller that the number of alive processes. This, however, implies that all processes get the same set of messages in round $r$, and thus they all have the same set of values in their respective values sets. Therefore the maximum of round $r$, denoted $m_r$ will be the same at all alive processes. Now agreement follows from the fact that no process can send a value $v \ne m_r$ in rounds $r' > r$, and therefore $\forall r' > r : m_{r'} = m_r$. Thus all processes will decide on $m_r$.

\section{Proofs for Algorithm $2$}

\subsection{Proof of Lemma \ref{lemma2.1}}

Since $p$ sends $(Lock, r, x, x)$ it cannot have received any propose messages for any other value (otherwise it could not have reached line $9$). Since processes wait for $\diamond \mathcal{N} \ge n - f > n/2$ messages, every other process must have received at least one $(Propose, r, x)$ message, keeping them from reaching line $9$ with $v \ne x$. Thus each processes either sends out a lock message for $x$ or $(Lock, r, ?, \_)$.

\subsection{Proof of Lemma \ref{lemma2.2}}

Let $p$ denote the deciding process; to be able to decide, $p$ requires $\diamond \mathcal{N} (Lock, r, x, \_)$ messages. From Lemma \ref{lemma2.1}, it follows that no process $q$ can have received sufficiently many messages to decide another value in this round. Therefore, (1) holds.

Since $p$ process has received $\diamond \mathcal{N} \ge n - f \ge f + 1$ lock messages, any other process must have received at least one of these and thus all reach line $20$ and calculate the same minimum, i.e., the only value $x$, and use this value as input for the next round. Since all alive processes now propose the same value in the proposal phase of round $r + 1$, all processes receive $\diamond \mathcal{N}$ messages containing $x$. This in turn results in all alive processes to lock $x$ and send enough lock messages to force all processed (that did not decide in $r$ and therefore terminated after broadcasting their lock messages) to decide in round $r + 1$, thereby ensuring (2).

\subsection{Proof of Lemma \ref{lemma2.3}}

For the sake of contradiction assume otherwise, and let $r_a$ denote the first round where the all alive processes¡¯ $\diamond \mathcal{N}$ is accurate at the start of the round (line $4$). Moreover, let $r_c$ denote the first round after the round in which the last process crashed. Since we assume that no process ever decides, both of these rounds must exist. Let $r_d = \max \{r_a, r_c\}$, then all processes will receive all the messages from all alive processes, in all rounds $r \ge r_d$. Therefore all must calculate the same minimum, say $x$, in line $7$ from $r_d$ on. If all values received were the same, this value is also locked by all processes and therefore decided on in the second phase, leading to a contradiction, since there is
a decision. So assume that there where different values in the propose messages, and thus $proposed - {x} \ne \emptyset$ ; in line $8$, which results in locked to contain $?$ in line $19$, i.e., no decision is possible in this round. However, all processes set $v \leftarrow x$ in line $27$. Now all processes have the same input value at the start of round $r_d + 1$, which leads to all processes to decide $x$ by the argument for (2) in Lemma \ref{lemma2.2}.

\subsection{Proof of Lemma \ref{lemma2.4}}

Since no process ever sets $v$ to a value that is neither its own initial value nor a value received from another process, it follows trivially that when $v$ is decided on, this value must
be some process¡¯s input value, thereby ensuring Validity. Finally, Integrity follows from halting in line $15$ in the round after deciding.

\section{Proofs for Algorithm $3$}

\subsection{Proof of Lemma \ref{lemma3.1}}

The proof is by contradiction. Let $r$ be the smallest round in which a process blocks forever. Blocking can only occur at one of the three wait statements. We will show that it is impossible to wait forever for each of them.

The first wait, requires that in phase $r$, no process will ever trust itself to be the leader, and thus not unblock itself directly and all other correct processes via a $(Leader, r, \_)$ message, contradicting the properties of $\Theta$.

Since no correct process can block forever at the first wait, all correct processes will eventually send a $(Report, r, \_)$ message thus unblocking all processes in the second wait (since
at most $f$ processes can crash and thus not send a $(Report, r, \_)$ message). The same is true about the third wait statement and $(Vote, r, \_)$ messages.

Therefore, no correct process was blocked forever in each of the three waits and thus they will all start round $r+1$, which contradicts the assumption that some processes will block in round $r$.

\subsection{Proof of Lemma \ref{lemma3.2}}

When one process decides, it has successfully broadcast a $(Decide, v)$ message in the line before. This message will eventually arrive at every other process and cause it to decide (if it did not decide before). Thus, when one process decides, every correct process decides.

We now prove that at least one process sends a $(Decide, v)$ message. Due to Lemma \ref{lemma3.1} and the properties of $\Theta$, eventually there is a round in which only one process sends a $(Leader, r, v)$ message, since this message is eventually received by all correct processes. They will all set their $v$ to the same value $w$, and thus $n - f$ identical messages will be sent and received in the second phase and thus $aux = w$ at all alive processes. And thus every alive process will send a $(Decide, v)$ message (with the $v = w$).

\subsection{Proof of Lemma \ref{lemma3.3}}

We show that when one of the two messages (w.l.o.g. $(Decide, w)$) is sent in round $r$, then (1) the other cannot be sent in $r$ or in later rounds, and (2) all processes have $v = w$.

First we note that in some given round $aux$ cannot take two different non-$?$ values, as only one value can reach a majority in a benign system. As the value sent out via $(Decide, w)$
messages cannot be $?$, it is clear that $w = aux$. Thus all $(Decide, \_)$ messages sent in $r$ must contain the same value.

Secondly, we observe that $v$ contains the same variable at all correct processes at the end of round $r$ since one process sending out a $n - f$ message in line $26$ implies that every
other process will receive at least one $(Vote, r, aux')$ with $aux' \ne ?$ since $n > 2f$. From this it follows that no other value can ever be decided after round $r$ (no other value will ever be proposed by a leader).

\section{Proof of Theorem \ref{diamond}}

Quite simply, eventually $\diamond \mathcal{N}$ will return the correct number of processes that have not crashed, let this time be denoted by $t'$. Eventually all messages from crashed processes have been delivered at all processes, let this time be denoted by $t''$. Let further $t = \max(t', t'')$ and $r_{max}$ the maximum round at time $t$. After $t$ we know that the output of $\diamond \mathcal{N}$ is accurate, thus all processes can only succeed to set their round to some value $r > r_{max} + 1$, if they have received $(Alive, r - 1)$ messages from all alive processes (which are the correct processes). These processes can be found in $AL$ at each correct process. Therefore from this time on, the set of suspected processes we have $suspect = P - AL = SF$, where $SF$ denotes the set of processes that have crashed. In other words $p$ will permanently suspect all crashed processes. Thus we have shown that $\diamond \mathcal{N} \diamond \mathcal{P}$ implements Strong Completeness. Moreover, since no correct process is suspected in any round that starts after $t$ it follows that Eventual Strong Accuracy is guaranteed as well.

\section{Proofs for Algorithm $5$}

\subsection{Proof of Lemma \ref{lemma4.1}}

Without loss of generality assume that round number of $q$ is further advanced than the round number of $p$. We now assume by contradiction that $q$ has reached round $r_p + f + 2$ (with $r_p$ denoting $p$¡¯s round number). Obviously $q$ did not receive a round $r'$ message from $p$ for any $r' > r_p$. The only way for $q$ to pass the wait statement in line $6$ is for some other process to send its round $r'$ message and then crash, thus eventually decreasing the output of $\mathcal{N}$ at $q$. Since $q$ has reached round $r_p + f + 2$ this must have happened $f + 1$ times, which contradicts the definition of f as the maximum number of failures in any execution.

\subsection{Proof of Lemma \ref{lemma4.2}}

Assume by contradiction that some process $p$ is put into the suspect set of some process $q$ before it crashes. This requires that $p$ is not in the alive processes set $AL$ of $q$ for $f + 2$ rounds. Moreover for $p$ to turn up in $q$'s suspect set the contents of $AL$ cannot have changed for $f + 2$ rounds. This also implies that $\mathcal{N}$ did not change for $f + 2$ rounds. Due to the accuracy property of $\mathcal{N}$ (it never outputs a number smaller than the number of alive processes) this in turn implies that no process crashed while $q$ performed the previous $f + 2$ rounds. Due to Lemma \ref{lemma4.1} no process that crashed before can have sent messages for all these rounds. Thus, $p$ must be in $AL$ contradicting the assumption that it was not.

\subsection{Proof of Lemma \ref{lemma4.3}}

Since a faulty process only takes finitely many steps, it can only send messages for finitely many rounds. Therefore, there exists a round from which on no process receives messages
from it, thus it is not part of the set $AL$ (the alive processes estimate) at any process. This results in crashed processes to be eventually suspected.

\end{appendix}

\end{document}